\input{aipcheck}

\documentclass[
    ,final            
  ]
  {aipproc}

\layoutstyle{6x9}

\usepackage{multirow}

\begin{document}

\title{Direct photon measurement in Au+Au collisions at $\sqrt{s_{NN}}$=200GeV at RHIC}

\classification{24.85.+p, 25.75.-q, 25.75.Nq, 25.75.Dw}
\keywords      {Quark Gluon Plasma, RHIC, Direct Photons, Dileptons, Au+Au}

\author{Takao Sakaguchi for the PHENIX Collaboration}{
  address={Brookhaven National Laboratory, Physics Department, Upton, NY 11973, U.S.A.}
}

\begin{abstract}
Direct photon production in Au+Au collisions at $\sqrt{s_{NN}}$=200\,GeV
has been measured. The result is compared to several theoretical calculations,
and found that it is not inconsistent with ones including thermal radiation
from QGP or jet-photon conversion process on top of a NLO pQCD expectation.
The direct photon contribution in dilepton measurement is also evaluated.
\end{abstract}

\maketitle


\section{Introduction}
Of many observables proposed for investigating Quark Gluon Plasma (QGP) that
is likely produced in relativistic heavy ion collisions, photons are
considered to be an excellent probe because they do not interact strongly
with medium once produced. The high $p_T$ direct photon measurement made
by the PHENIX experiment~\cite{bib1} at RHIC provided a strong evidence
that the initial hard scattering cross-section is not suppressed, and the jet
quenching is due to a final state interaction (energy loss)~\cite{bib2}.
There are possible contributions of thermal radiation from QGP in
1$<$$p_T$$<$2.5\,GeV/$c$~\cite{bib3} and jet-photon conversion in
$p_T$$>$2.5\,GeV/$c$~\cite{bib4}, but they have not been observed because of a
large systematic error. This paper presents the latest measurement on direct
photons in Au+Au collisions at $\sqrt{s_{NN}}$=200\,GeV from RHIC 2004 run.

\section{Results and Discussions}
\begin{figure}
\centering
\begin{tabular}{lll}
\begin{minipage}{35mm}
\includegraphics[width=3.3cm]{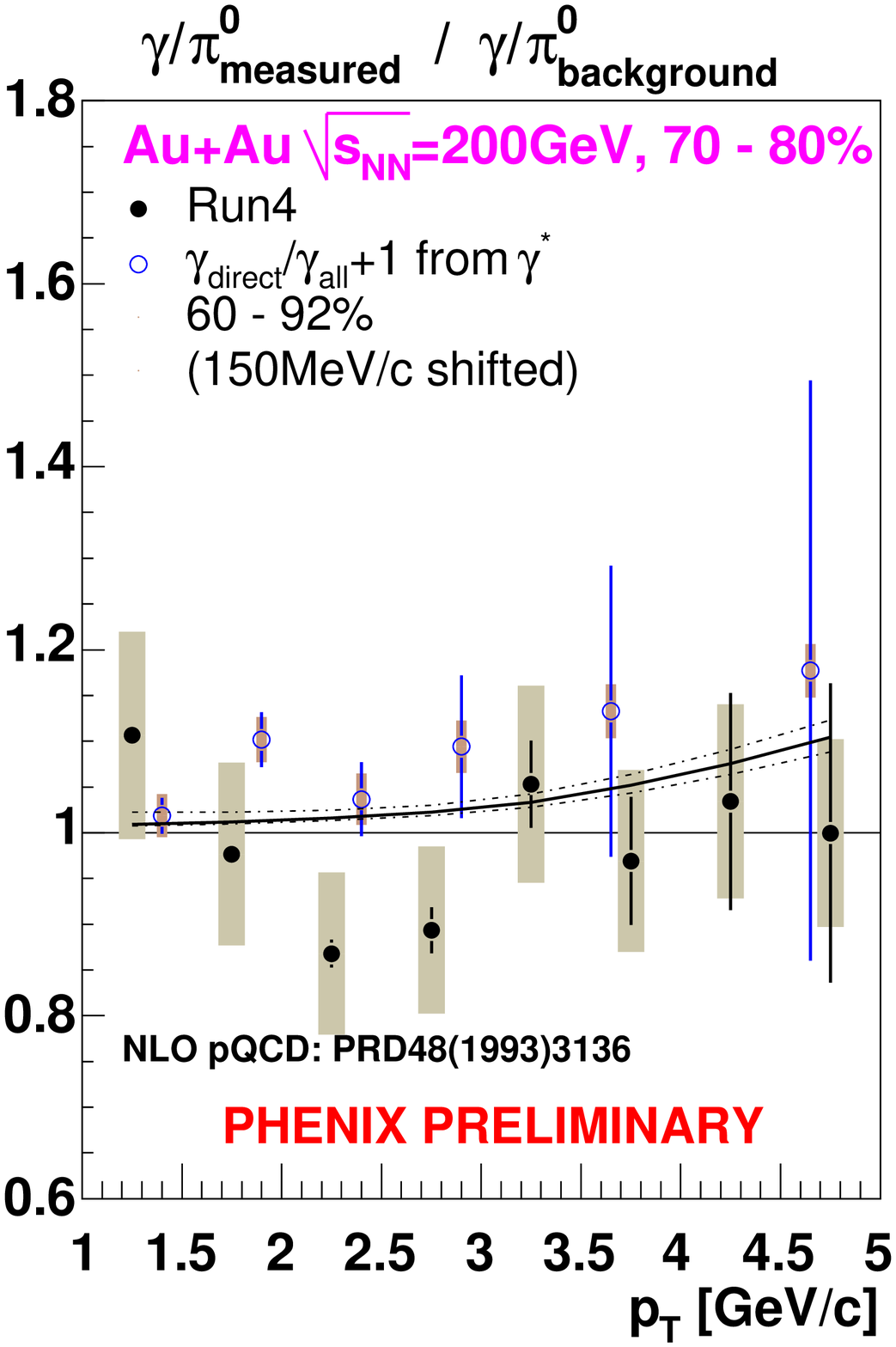}
\end{minipage}
&
\begin{minipage}{35mm}
\includegraphics[width=3.3cm]{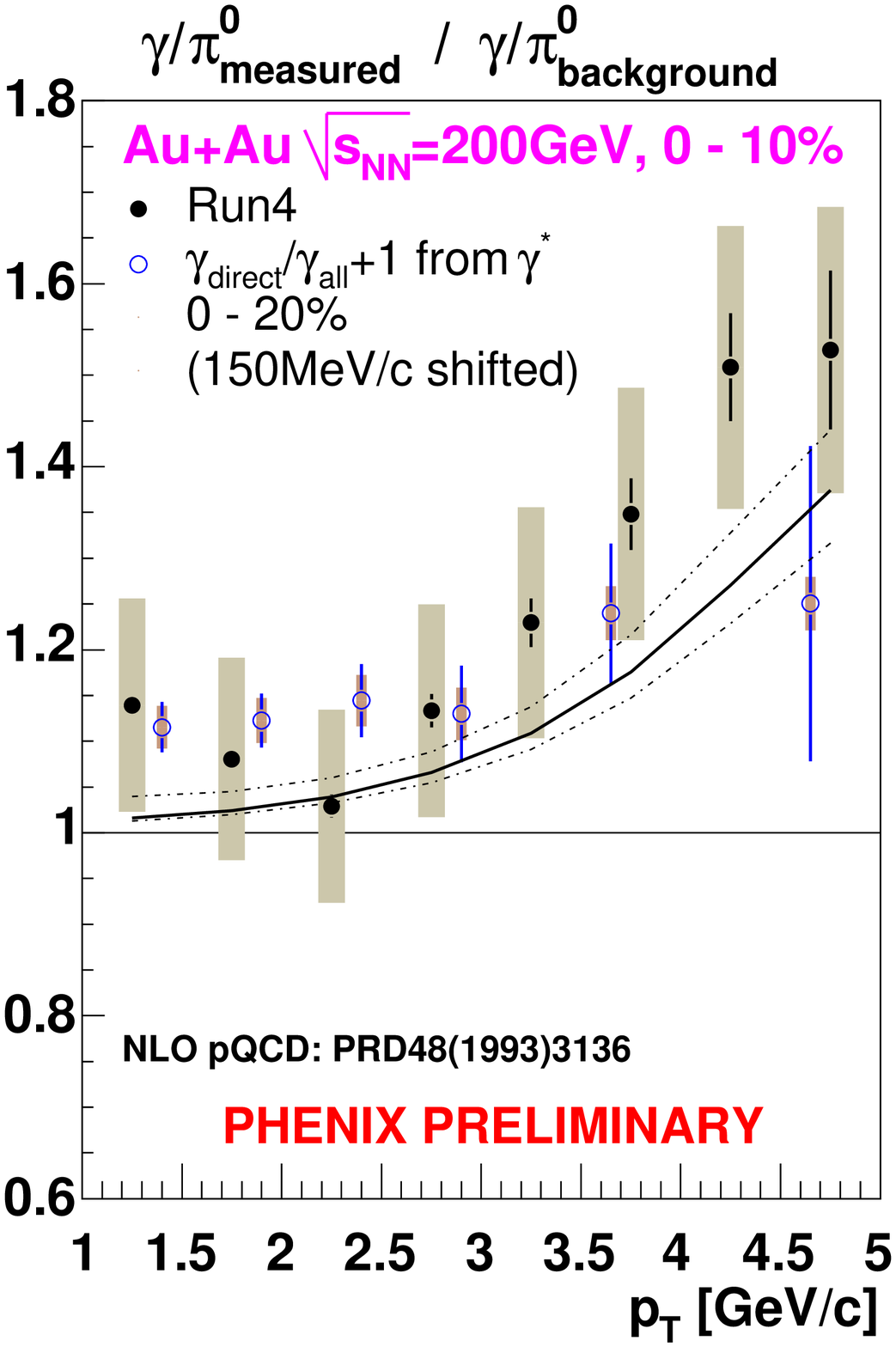}
\end{minipage}
&
\hspace{-5mm}
\multirow{2}{67mm}[17mm]{
\begin{minipage}{67mm}
\includegraphics[width=6.7cm]{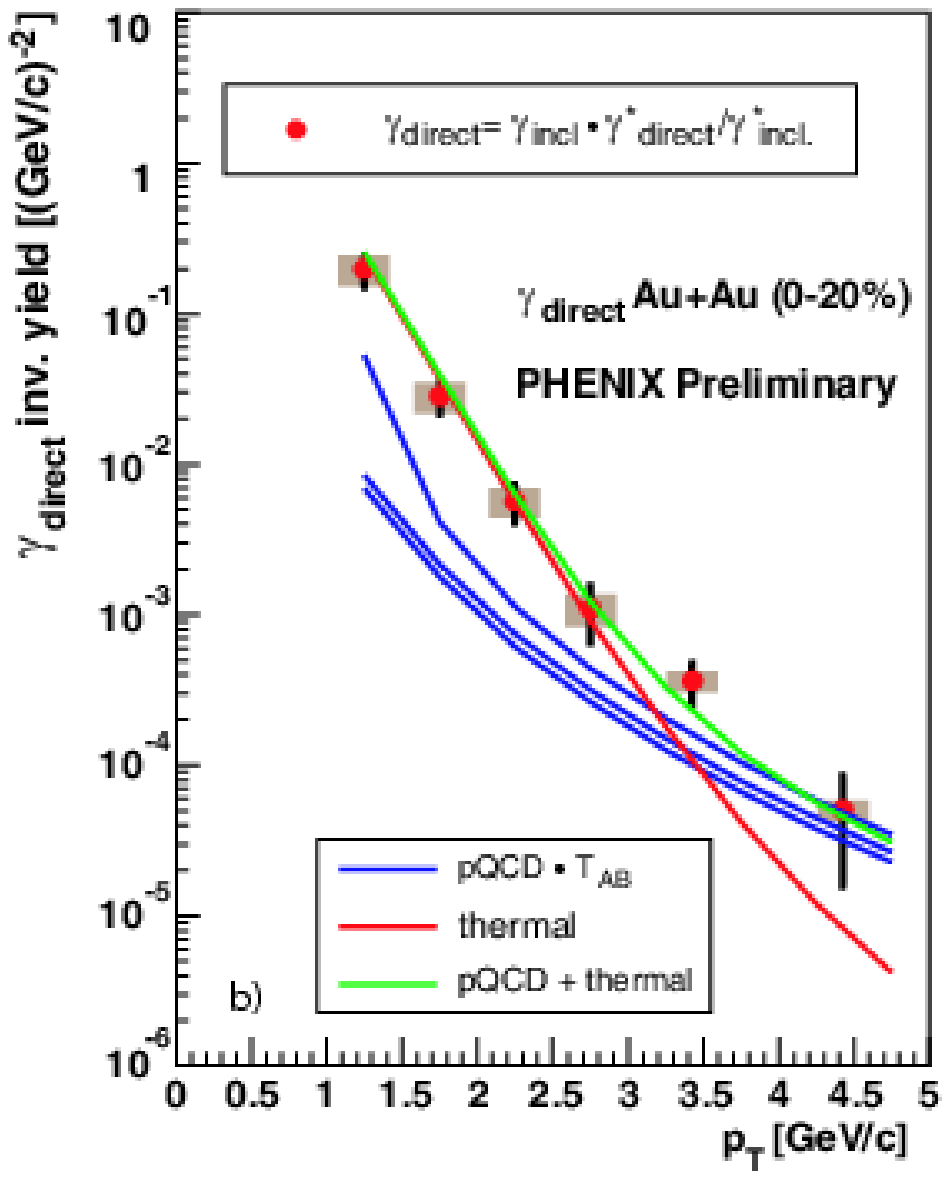}
\end{minipage}
}
\\
\begin{minipage}{35mm}
\includegraphics[width=3.3cm]{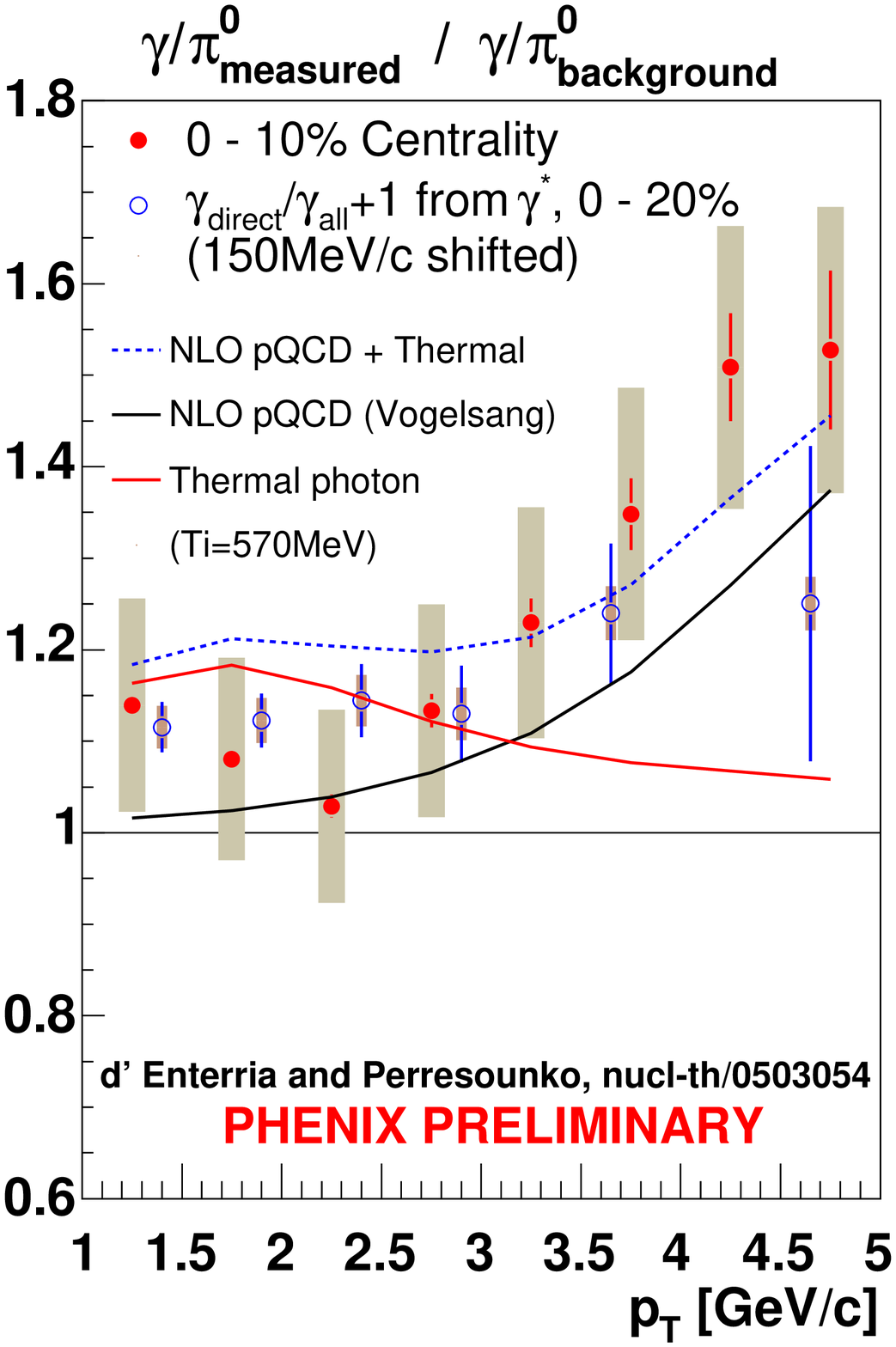}
\end{minipage}
&
\begin{minipage}{35mm}
\includegraphics[width=3.3cm]{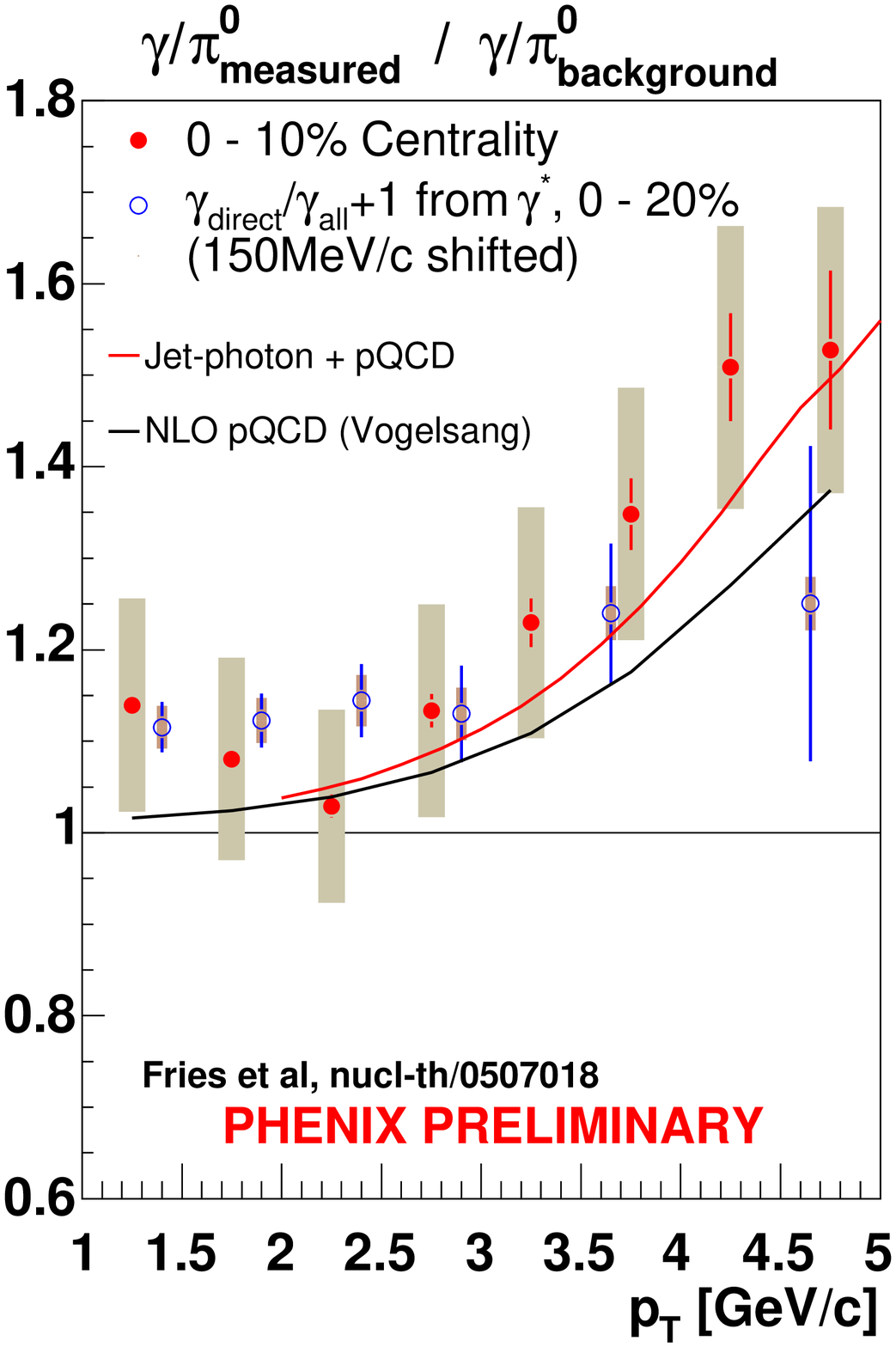}
\end{minipage}
& \\
\end{tabular}
\caption{Double ratios for direct photons for (a) 70-80\,\% (top left) and (b) 0-10\,\% (top middle) centrality, together with $\gamma^*_{direct}/\gamma^*_{all} +1$ from very low mass dilepton analysis. A NLO pQCD expectation scaled by the number of binary collisions are shown in lines. (c) Direct photon spectrum for 0-20\,\% is obtained from $\gamma^*$ result (right). The 0-10\,\% result is compared to models including (d) thermal radiation (bottom left), and (e) jet-photon conversion (bottom middle).}
\label{fig1}
\end{figure}

There are two analyses for extracting direct photon signal. One is to estimate
background real photon spectra from known hadronic sources and take ratios to
measured inclusive photon spectra to look for an excess. The other is to look
at low mass dileptons ($\gamma^*$, 90$<$$M_{ee}$$<$300\,MeV/$c^2$), and follow
almost the same procedure as the real photon analysis. In the latter analysis,
the excess ratio is converted into the one at the very low mass dilepton
(0$<$$M_{ee}$$<$30\,MeV/$c^2$), which is approximately equal to the one of
total yields. Detail description can be found in the literature~\cite{bib5}.
Figures~\ref{fig1}(a) and (b) show
$(\gamma/\pi^0_{measured})/(\gamma/\pi^0_{background})$ (double ratio) for the
real photon measurement for 0-10\,\% and 70-80\,\% centrality, together with
the very low mass dilepton analysis results.
\renewcommand{\multirowsetup}{\centering}
They agree each other given a large systematic error of real photon
measurement. The NLO pQCD expectation~\cite{bib7} scaled by number of binary
collisions are shown as well. They are also consistent with already published
results from Run2 analysis~\cite{bib2}. The direct photon spectrum for
0-20\,\% central collisions shown in Fig.~\ref{fig1}(c) was obtained by
$\gamma_{dir}$=$\gamma^*_{direct}/\gamma^*_{all}\times\gamma_{inclusive}$.
Overlaid is a theoretical prediction including thermal radiation from
QGP~\cite{bib8}.
Figs.~\ref{fig1}(d) and (e) show comparisons of the 0-10\,\% central results
with theoretical models. The result is consistent with models including
thermal radiation and jet-photon conversion process. However, the recent
measurement in p+p collisions shows a trend that the data are higher than
the NLO expectation for $p_T$$<$5\,GeV/$c$ though the systematic errors
cover that deviation~\cite{bib9}. The fact can potentially wash out the
contributions discussed here. Therefore, a further work on reducing
systematic errors in real photon measurement both in Au+Au and
p+p is needed to make a concrete statement.

\begin{figure}
\centering
\begin{minipage}{60mm}
\includegraphics[width=5.5cm]{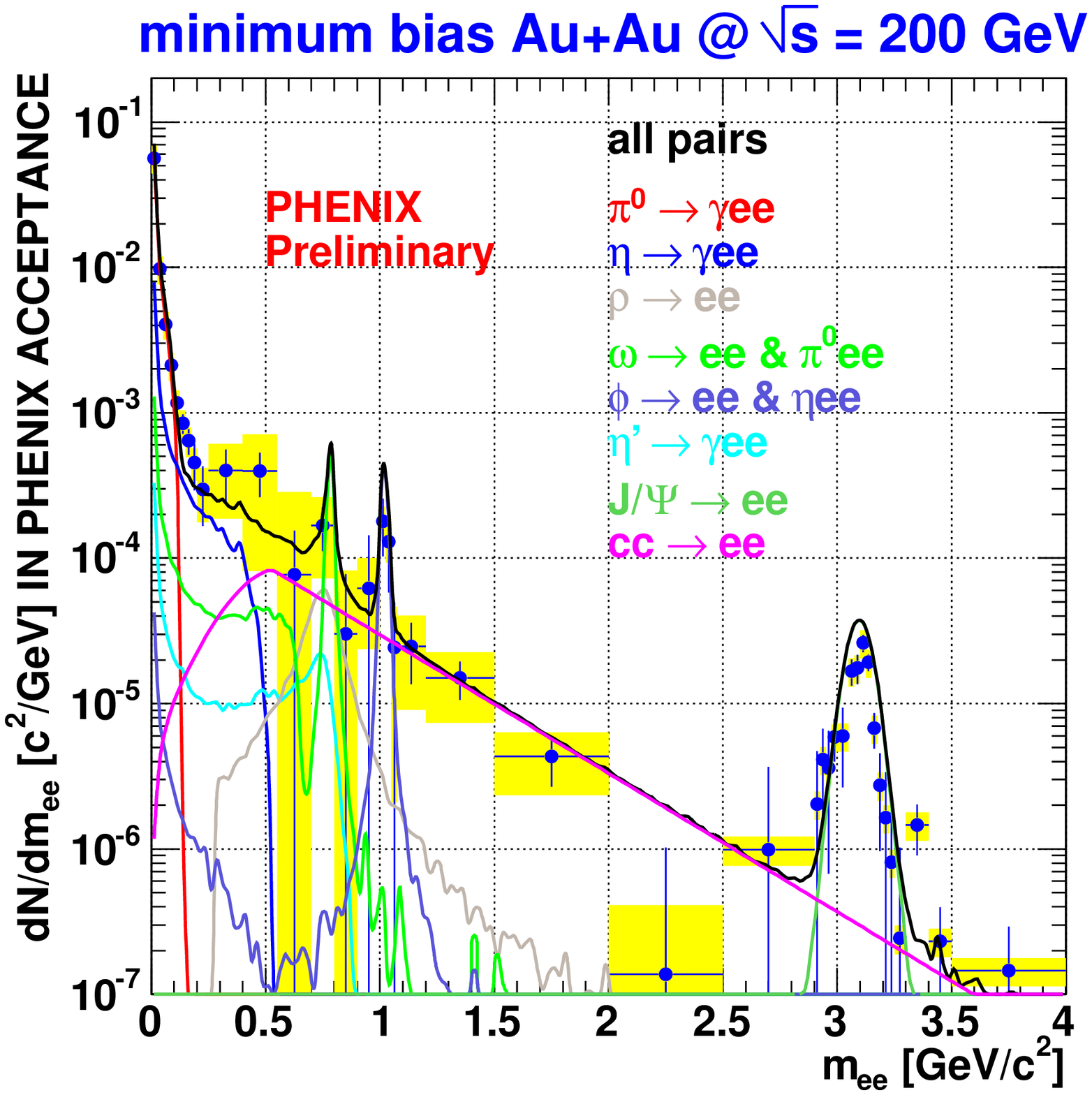}
\end{minipage}
\hspace{10mm}
\begin{minipage}{60mm}
\includegraphics[width=5.5cm]{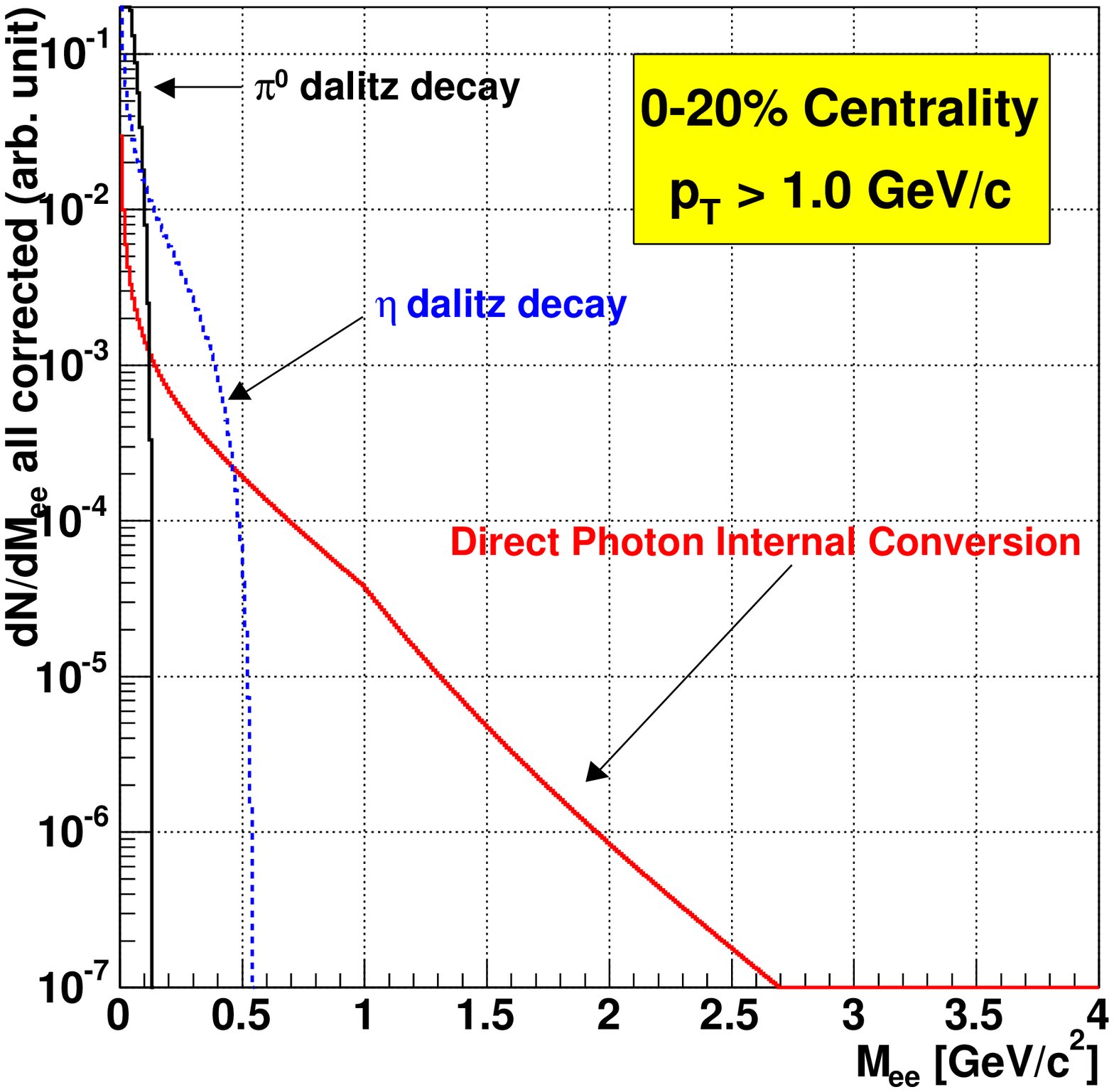}
\end{minipage}
\caption{(a) Dilepton spectrum for Au+Au minimum bias events (left), and (b) simple monte carlo calculation on contribution of direct photons in dileptons (right).}
\label{fig3}
\end{figure}

The thermal radiation from QGP can also be seen with dileptons. They are
produced through annihillation of quarks and anti-quarks, which is a part
of the same processes of producing direct photons. Fig.~\ref{fig3}(a) shows
the dilepton spectrum for minimum bias events~\cite{bib10}. The data is
corrected for efficiencies, but neither for acceptance nor momentum smearing.
The acceptance and momentum smearing are taken into account in $e^+e^-$
cocktail calculation shown as lines. There is no excess seen over the
contributions from known background sources within the systematic and
statistic errors over $M_{ee}$ regions.
The internal conversion of direct photons should contribute to dilepton
spectra not only in very low mass region, but also high mass region.
In order to look for the contributions of direct photon internal conversion,
a simple monte carlo is carried out. Fig.~\ref{fig3}(b) shows the $e^+e^-$
invariant mass distribution for 0-20\,\% centrality for $p_T$$>$1.0\,GeV/$c$
estimated using measured $\pi^0$, $\eta$ and direct photon spectra using
Kroll-Wada formula~\cite{bib6}. Although Figs.~\ref{fig3}(a) and (b) are under
different conditions on centrality selection, $p_T$ selection, and neither
filtered in the detector acceptance nor efficiency, the contribution of direct
photon internal conversion in the higher
$M_{ee}$ region is found to be negligibly small. It is also confirmed that
the direct photon result from the very low mass dilepton analysis is
consistent with the observed dilepton result.

\section{Conclusion}
Direct photon production in Au+Au collisions at $\sqrt{s_{NN}}$=200\,GeV
has been measured. The direct photon yield is in agreement with published
data from RHIC Year-2 Run, and the one from a method that explore very low
mass dilepton. The result is compared to several theoretical calculations,
and found that it is not inconsistent with ones including thermal radiation
from QGP or jet-photon conversion process on top of a NLO pQCD expectation.
The direct photon contribution in dilepton measurement is evaluated, and
found it is negligible at high $M_{ee}$.
A further work on reducing systematic errors on direct photon measurement
both in Au+Au and p+p is desired to make a concrete statement.

\bibliographystyle{aipproc}   

\end{document}